# Multiferroicity in an organic charge-transfer salt: Electric-dipole-driven magnetism

P. Lunkenheimer[1]*, J. Müller[2], S. Krohns[1], F. Schrettle[1], A. Loidl[1], B. Hartmann[2], R. Rommel[2], M. de Souza[2,#], C. Hotta[3], J.A. Schlueter[4], M. Lang[2]

**Multiferroics, showing simultaneous ordering of electrical and magnetic degrees of freedom, are remarkable materials as seen from both the academic and technological points of view[1,2]. A prominent mechanism of multiferroicity is the spin-driven ferroelectricity, often found in frustrated antiferromagnets with helical spin order[1,3,4,5]. There, similar to conventional ferroelectrics, the electrical dipoles arise from an off-centre displacement of ions. However, recently a different mechanism, namely purely electronic ferroelectricity, where charge order breaks inversion symmetry, has attracted considerable interest[6]. Here we provide evidence for this exotic type of ferroelectricity, accompanied by antiferromagnetic spin order, in a two-dimensional organic charge-transfer salt, thus representing a new class of multiferroics. Quite unexpectedly for electronic ferroelectrics, dipolar and spin order arise nearly simultaneously. This can be ascribed to the loss of spin frustration induced by the ferroelectric ordering. Hence, here the spin order is driven by the ferroelectricity, in marked contrast to the spin-driven ferroelectricity in helical magnets.**

In the present work, we have investigated single crystalline κ-(BEDT-TTF)$_2$Cu[N(CN)$_2$]Cl (κ-Cl), where BEDT-TTF stands for bis(ethylenedithio)-tetrathiafulvalene (often abbreviated as ET). Two crystals with different geometries and contact materials were investigated (see methods section). In these compounds, dimers of ET molecules form an anisotropic triangular lattice with a half-filled dimer band, where the strong on-dimer Coulomb interaction $U$ drives the system to a Mott insulating state[7,8]. In addition, the importance of intra-dimer degrees of freedom and inter-site interactions $V$ have been pointed out[9,10,11]. κ-Cl consists of alternating conducting ET layers and insulating anion sheets (see Supplementary Information (SI), Fig. S1). Within the ET layers, adjacent molecules form dimers on which a single electron hole is located. Below $T_N \approx 27$ K, intralayer antiferromagnetic and interlayer ferromagnetic ordering of hole spins occur, followed by weak ferromagnetic canting below 23 K (refs. 12,13). κ-Cl becomes superconducting below 12.8 K, when applying pressures of 300 bar[14].

Figure 1 shows the temperature dependence of the conductivity σ' of crystal 1, measured at 2.1 Hz, providing a good estimate of the dc conductivity $\sigma_{dc}$, see SI. Aside of the well-known overall semiconducting characteristics of $\sigma_{dc}(T)$[14,15], we find a jump-like decrease by two decades at around 27 K, about the same temperature where long-range antiferromagnetic ordering is reported[7,8]. A corresponding jump was also found in sample 2 (see SI). A distinct drop in $\sigma_{dc}$ has not been observed in most ac and dc studies reported so far (except in ref. 16), which may be ascribed to the higher precision of dielectric measurements at high resistances, compared to standard dc setups. This abrupt reduction of the conductivity within the Mott insulating state is a first hint to a charge-order (CO) transition, i.e. a further localization of charge carriers.

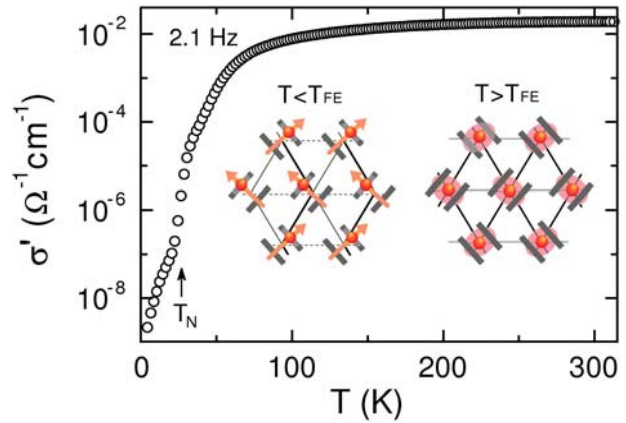

**Figure 1 | Temperature dependence of conductivity of κ-(ET)$_2$Cu[N(CN)$_2$]Cl.** Conductivity of sample 1 at 2.1 Hz for **E**∥b ($b$ denotes the interlayer direction), representing a reasonable estimate of the dc conductivity. Arrow indicates the Néel temperature reported in ref. 12, see SI for sample-to-sample variations. The insets show schematic drawings of the ac planes for temperatures below and above $T_{FE}$. Thick grey lines: ET molecules, red spheres: holes (the red shaded areas for $T > T_{FE}$ indicate their delocalization), orange arrows: dipolar moments arising at $T < T_{FE}$.

Figure 2 shows the dielectric constant $\varepsilon'(T)$ of sample 1 for various frequencies. Pronounced peaks reaching absolute values up to several hundreds are revealed. While the peak positions are nearly frequency independent, their amplitudes become strongly suppressed with increasing frequency. The overall behaviour is typical for order-disorder-type ferroelectrics, where the electric dipoles that are disordered at high temperatures order with net overall polarization below a phase-transition temperature $T_{FE}$ (refs. 17,18). Identifying the dipoles with the holes fluctuating within the dimers at $T > T_{FE}$ and cooperatively locking in to one of the two ET molecules at $T < T_{FE}$, provides a plausible scenario for the present material (see insets of Fig. 1). Similar behaviour as observed in Fig. 2 was also found for other ferroelectric charge-transfer salts[19,20]. The peaks are superposed to a rather high background contribution $\varepsilon_b$ of about 120, partly arising from stray-capacitance contributions due to the small sample dimensions. The dashed line in Fig. 2 demonstrates that the high-temperature flanks of $\varepsilon'(T)$ at low frequencies (i.e., the static dielectric constant $\varepsilon_s(T)$) are consistent with a Curie-Weiss behaviour with an additional background term, $\varepsilon_s = C/(T-T_{CW}) + \varepsilon_b$, with a Curie-Weiss temperature $T_{CW} = 25\,K \approx T_{FE}$. Sample 2 showed similar dielectric response to that seen in Fig. 2 (see SI), despite having significantly different geometry and when different contact material was used – clear evidence for the intrinsic nature of the observed behaviour. In a previous dielectric investigation of κ-Cl[21], relaxational behaviour was deduced from frequency-dependent dielectric measurements below 30 K, consistent with the notion of order-disorder ferroelectricity[17,18]. Due to the restricted temperature range in that work, no $\varepsilon'(T)$-peaks were detected[21]. Likewise, measurements of the in-plane dielectric properties showed no signatures of ferroelectric order.[22]

---

[1]Experimental Physics V, Center for Electronic Correlations and Magnetism, University of Augsburg, 86159 Augsburg, Germany
[2]Institute of Physics, Goethe-University Frankfurt, Max-von-Laue-Str. 1, 60438 Frankfurt(M), Germany
[3]Department of Physics, Faculty of Science, Kyoto Sangyo University, Kyoto 603-8555, Japan
[4]Materials Science Division, Argonne National Laboratory, Argonne, IL 60439, USA
#Present address: Departamento de Física, IGCE, Unesp - Universidade Estadual Paulista, Caixa Postal 178, CEP 13500-970, Rio Claro (SP), Brazil
*e-mail: peter.lunkenheimer@physik.uni-augsburg.de

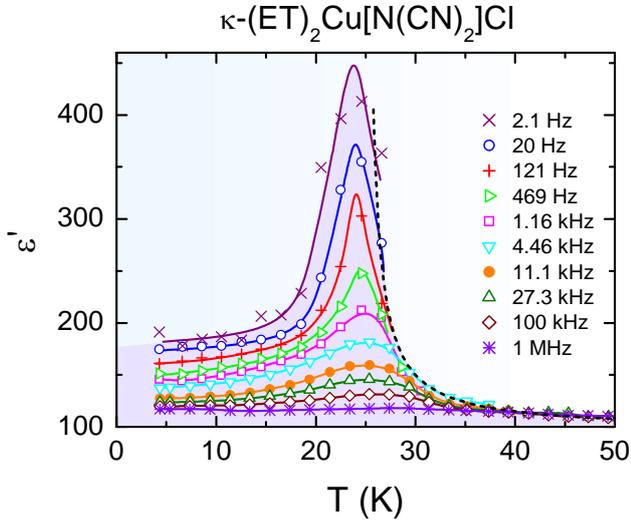

**Figure 2 | Temperature-dependence of the dielectric constant.** $\varepsilon'(T)$ for sample 1, for selected frequencies and $E \| b$. The solid lines are guides for the eyes. The dashed line demonstrates Curie-Weiss behaviour with $T_{CW} = 25$ K.

Further evidence for ferroelectric ordering in κ-Cl is provided by so-called "positive-up-negative-down" (PUND) measurements[23], where a sequence of trapezoid pulses (Fig. 3a) is applied to the sample. The resulting current response (Fig. 3b) shows strong peak-like features occurring when the electric field $|E|$ exceeds about 10 kV/cm for the first and third pulse. Obviously, here the macroscopic polarization of the sample is switched, i.e. the dipoles within the ferroelectric domains reorient along the field, implying a motion of charges and, thus, a peak in $I(t)$. As expected for ferroelectrics, at the second and forth pulse no corresponding features show up since the polarization was already switched by the preceding pulse. (The horizontal sections in $I(t)$ represent the trivial charging and discharging currents of the sample capacitor.) The typical PUND response was clearest for low frequencies, $\nu < 100$ Hz.

Stimulated by these results, we have also performed field-dependent polarization measurements, $P(E)$. Typical results are shown in Fig. 3c. Above $T_{FE}$, elliptical curves are observed (crosses), typical for linear polarization response (paraelectricity) with additional loss contributions from charge transport[18]. However, at lower temperatures, a clear onset of non-linearity above about 8.5 kV/cm occurs and saturation at the highest fields is observed (circles), as expected for ferroelectrics. The saturation polarisation of several tenths of µC/cm² compares well to that of the ferroelectric charge-transfer salt tetrathiafulvalene-p-bromanil (0.2 µC/cm²)[24] and the electronic ferroelectric and multiferroic magnetite (0.5 µC/cm²)[6]. In the electronic ferroelectrics LuFe₂O₄ (25 µC/cm²)[25] and $Pr_{1-x}Ca_xMnO_3$ (44 µC/cm²)[26], somewhat higher values are found. Interestingly, at higher fields, $|E| > 12$ kV/cm, breakdown effects were observed. This may indicate that the sample resistance has become too low and that interplane charge transport sets in. Thus there is a rather restricted range of roughly 9-12 kV/cm for nonlinear polarization response in κ-Cl.

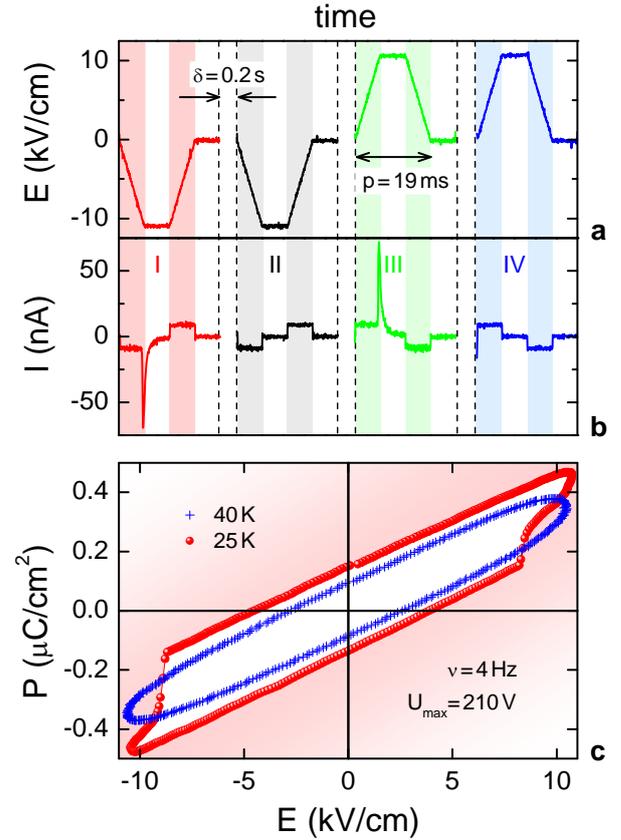

**Figure 3 | Electric polarization switching. a,** Time-dependent excitation signal $E\|b$ of the PUND measurements performed at 25 K with waiting time $\delta$ and pulse width $p$. **b,** Resulting time-dependent current of sample 2. The spikes in response to pulses I and III provide evidence for polarization switching. **c,** Polarization-field hysteresis curves at temperatures below and above the ferroelectric phase transition.

Overall, the results presented in Figs. 2 and 3 provide strong evidence for ferroelectricity in κ-Cl, coinciding with the occurrence of magnetic ordering. The appearance of simultaneous electrical and magnetic order is a well understood phenomenon in multiferroic systems with spiral spin order[1,2,3]. There the onset of helical spin order breaks inversion symmetry and, due to the Dzyaloshinskii-Moriya (DM) interaction, triggers the formation of displacive ferroelectric order via the electro-magnetic coupling $\mathbf{P} \propto \mathbf{e} \times \mathbf{Q}$. Here $\mathbf{Q}$ denotes the propagation vector of spin order and $\mathbf{e} = \mathbf{S_i} \times \mathbf{S_j}$, a cross product between adjacent spins $\mathbf{S_i}$ and $\mathbf{S_j}$, gives the spiral axis[2,4]. Thus, these systems are regarded as spin-driven ferroelectrics. Could this be the mechanism generating multiferroicity in κ-Cl? Indeed, it has been pointed out that a finite DM interaction is present in κ-Cl[8,13]. However, to our knowledge no experimental evidence for a helical magnetic structure in this material has been found until now.

To check for a spin-driven mechanism, we have performed magnetic field-dependent dielectric measurements. Usually, spin-driven multiferroics show a strong dependence of the dielectric properties on externally applied magnetic fields (so-called magnetocapacitive effects)[1,2,27]. Moreover, in κ-Cl a spin-flop transition at 0.25 Tesla into a direction perpendicular to the field has been found below 23 K (refs. 12,13). In the spin-driven scenario, the spin-flop transition would affect the dielectric properties through $\mathbf{P} \propto \mathbf{Q} \times (\mathbf{S_i} \times \mathbf{S_j})$. Remarkably enough, no variation of the dielectric properties in fields up to 9 T was found (see SI), ruling out a spin-driven mechanism for multiferroicity in κ-Cl.

This suggests that for κ-Cl a purely electronic, CO-driven ferroelectric state, as also considered for other ferroelectric charge-transfer salts[6,9,19,28], is realized. In fact, in the seminal article by van den Brink and Khomskii[6], charge-transfer salts were considered as good candidates for multiferroics with a ferroelectric state driven by CO. The mechanism discussed in that work is based on a combination of bond- and site-centred CO leading to the breaking of inversion symmetry. This provides a plausible scenario for κ-Cl: Just as in the structurally related κ-(ET)$_2$Cu$_2$(CN)$_3$, at high temperatures each hole can be assumed to be delocalized on an ET dimer[9,10,11]. This corresponds to pure bond centring (cf. Fig. 1c of ref. 6). Below $T_{FE}$, the holes lock in to one of the two molecules of the dimers, establishing additional site centring (Fig. 1d in ref. 6). This will primarily lead to a polarization parallel to the planes, and, due to the inclined spatial orientation of the ET molecules[8], a polarization along $b$ is also induced.

In contrast to the spin-driven systems, for CO multiferroics, the ferroelectric and magnetic transition temperatures can differ significantly[6,25,29]. How can the nearly identical transition temperatures in κ-Cl be explained within the above framework? The magnetism in κ-Cl arises from the holes' spins, which, at $T > T_N$, are already well defined on each dimer. As the holes fluctuate within the dimers, their spins – on average – sit at the centres of the dimers, forming an anisotropic triangular lattice within the $ac$ planes (indicated by lines in the schematic drawing in the inset of Fig. 1), giving rise to geometrical frustration. Indeed, in κ-(ET)$_2$Cu$_2$(CN)$_3$, where frustration is even stronger, a spin-liquid state is formed[30]. There, dielectric measurements[9] have revealed relaxor ferroelectricity, i.e. short-range cluster-like ferroelectric order. However, in κ-Cl we find long-range ferroelectricity, caused by CO, which implies the collective off-centre positioning of holes within the dimers below $T_{FE}$. Consequently, the holes (and spins) no longer reside on a highly-frustrated triangular lattice (inset of Fig. 1). The frustration should therefore be reduced and magnetic order sets in (see SI).

In summary, the observed conductivity, dielectric and polarization properties of κ-Cl provide evidence for CO-driven electronic ferroelectricity, making this material the first example of a multiferroic charge-transfer salt. Moreover, our measurements indicate that ferroelectric order drives the magnetic order, in contrast to the well-known spin-driven mechanism in the helical magnets! Thus κ-Cl not only represents a new class of multiferroics but also provides the first example for a new coupling mechanism of magnetic and ferroelectric ordering.

## Methods

**Crystal growth.** Crystals of κ-(BEDT-TTF)$_2$Cu[N(CN)$_2$]Cl were grown according to literature methods.[15,31] Sample 2 was grown on the surface of a platinum electrode, as is typical for growth by electrocrystallization. It was roughly plate-like ($b$-axis vertical to the surface) with an area of about $0.3 \times 0.5$ mm$^2$ and a thickness of 0.2 mm. Occasionally during the electrocrystallization process, crystals grow on the bottom of the electrochemical cell, removed from the electric field at the electrode surface, and have a rod-like morphology. X-ray diffraction measurements show both morphologies are the same crystalline phase. Sample 1 was grown by this method. The long crystal axis (parallel to the $b$ crystallographic axis) was about 0.4 mm and a cross section of about 0.07 mm$^2$.

**Dielectric measurements.** For the dielectric measurements, contacts of graphite paste (sample 1) or evaporated gold (sample 2) were applied at opposite sides of the crystals, ensuring an electrical field direction parallel to $b$, i.e. in all measurements, the electrical field was directed perpendicular to the ET-planes. The dielectric constant and conductivity were determined using a frequency-response analyzer (Novocontrol α-Analyzer) and an autobalance bridge (Hewlett-Packard HP4284). For sample cooling at zero magnetic field, a $^4$He-bath cryostat (Cryovac) was used. Magnetic field-dependent dielectric measurements with H||$b$ up to 9 T were performed in a Quantum Design physical properties measurement system (PPMS). Additional measurements up to 1.5 T, with the magnetic field directed parallel to the $ac$ plane, were performed using a conventional electromagnet. In this case, sample cooling was achieved by a closed-cycle refrigerator. For the non-linear measurements, a ferroelectric analyzer (aixACCT TF2000) was used where cooling was also being provided by a closed-cycle refrigerator.

## Acknowledgements
We thank H.-A. Krug von Nidda and Harald Jeschke for helpful discussions. This work was supported by the Deutsche Forschungsgemeinschaft via the Transregional Collaborative Research Centers TRR 80 and TRR 49. Work at Argonne was supported by the U.S. Department of Energy Office of Science, operated under Contract No. DE-AC02-06CH11357.

## References

1. Fiebig M. Revival of the magnetoelectric effect. *J. Phys. D: Appl. Phys.* **38**, R123-R152 (2005).
2. Cheong S.-W. & Mostovoy M. Multiferroics: a magnetic twist for ferroelectricity. *Nat. Mater.* **6**, 13-20 (2007).
3. Sergienko, I. A. & Dagotto, E. Role of the Dzyaloshinskii-Moriya interaction in multiferroic perovskites. *Phys. Rev. B* **73**, 094434 (2006).
4. Katsura, H., Nagaosa, N. & Balatsky, A. V. Spin Current and Magnetoelectric Effect in Noncollinear Magnets. *Phys. Rev. Lett.* **95**, 057205 (2005).
5. Mostovoy, M. Ferroelectricity in spiral magnets. *Phys. Rev. Lett.* **96**, 067601 (2006).
6. van den Brink, J. & Khomskii, D. I. Multiferroicity due to charge ordering. *J. Phys.: Condens. Matter* **20**, 434217 (2008).
7. Kanoda, K. in *The Physics of Organic Superconductors and Conductors* (ed. Lebed, A.) (Springer, Berlin, 2008).
8. Toyota, N., Lang, M., & Müller, J. *Low-Dimensional Molecular Metals* (Springer, Berlin, 2007).
9. Abdel-Jawad, M. *et al.* Anomalous dielectric response in the dimer Mott insulator κ-(BEDT-TTF)$_2$Cu$_2$(CN)$_3$. *Phys. Rev. B* **82**, 125119 (2010).
10. Hotta, C. Quantum electric dipoles in spin-liquid dimer Mott insulator κ-(ET)$_2$Cu$_2$(CN)$_3$. *Phys. Rev. B* **82**, (R)241104 (2010).
11. Li, H., Clay, R. T. & Mazumdar, S. The paired-electron crystal in the two-dimensional frustrated quarter-filled band. *J. Phys.: Condens. Matter* **22**, 272201 (2010).
12. Miyagawa, K., Kawamoto, A., Nakazawa, Y. & Kanoda, K. Antiferromagnetic ordering and spin structure in the organic conductor, κ-(BEDT-TTF)$_2$Cu[N(CN)$_2$]Cl. *Phys. Rev. Lett.* **75**, 1174-1177 (1995).

13. Smith, D. F. *et al*. Dzialoshinskii-Moriya interaction in the organic superconductor κ-(BEDT-TTF)$_2$Cu[N(CN)$_2$]Cl. *Phys. Rev. B* **68**, 024512 (2003).
14. Wang, H. H. *et al*. New κ-phase materials, κ-(ET)$_2$Cu[N(CN)$_2$]X, X = Cl, Br and I - The sythesis, structure and superconductivity above 11 K in the Cl (T$_c$ = 12.8 K, 0.3 kbar) and Br (T$_c$ = 11.6 K) salts. *Synth. Met.* **41-43**, 1983-1990 (1991).
15. Williams J. M. *et al*. From semiconductor-semiconductor transition (42 K) to the highest-$T_c$ organic superconductor, κ-(ET)$_2$Cu[N(CN)$_2$]Cl (T$_c$ = 12.5 K). *Inorg. Chem.* **29**, 3272-3274 (1990).
16. Dressel, M. *et al*. Studies of the microwave resistivity of κ-(BEDT-TTF)$_2$Cu[N(CN)$_2$]Cl. *Synth. Met.* **70**, 927-928 (1995).
17. Blinc R. & Žekš, B. *Soft modes in ferroelectrics and antiferroelectrics* (North-Holland, Amsterdam, 1974).
18. Lines, M. E. & Glass, A. M. *Principles and application of ferroelectrics and related materials* (Clarendon Press, Oxford, 1977).
19. Nad, F. & Monceau, P. Dielectric response of the charge ordered state in quasi-one-dimensional organic conductors. *J. Phys. Soc. Japan*, **75**, 051005 (2006).
20. Starešinić, D., Biljaković, K., Lunkenheimer, P. & Loidl A. Slowing down of the relaxational dynamics at the ferroelectric phase transition in one-dimensional (TMTTF)$_2$AsF$_6$. *Solid State Commun*. **137**, 241-245 (2006).
21. Pinterić, M. *et al*. Magnetic anisotropy and low-frequency dielectric response of weak ferromagnetic phase in κ-(BEDT-TTF)$_2$Cu[N(CN)$_2$]Cl, where BEDT-TTF is Bis(ethylenedithio)tetrathiafulvalene. *Eur. Phys. J. B* **11**, 217-225 (1999).
22. Takahide, Y. *et al*. Highly nonlinear current-voltage characteristics of the organic Mott insulator κ-(BEDT-TTF)$_2$Cu[N(CN)$_2$]Cl. *Phys. Rev. B* **84**, 035129 (2011).
23. Scott, J. F. *Ferroelectric memories* (Springer, Berlin, 2000).
24. Kagawa, F., Horiuchi, S., Tokunaga, M., Fujioka, J. & Tokura, Y. Ferroelectricity in a one-dimensional organic quantum magnet. *Nature Phys*. **6**, 169-172 (2010).
25. Ikeda, N. *et al*. Ferroelectricity from iron valence ordering in the charge-frustrated system LuFe$_2$O$_4$. *Nature* **436**, 1136-1138 (2005).
26. Joos, Ch. *et al*. Polaron melting and ordering as key mechanisms for colossal resistance effects in manganites. *Proc. Natl. Acad. Sci. USA* **104**, 13597-13602 (2007).
27. Schrettle, F. *et al*. Switching the Ferroelectric Polarization by External Magnetic Fields in the Spin = 1/2 Chain Cuprate LiCuVO$_4$. *Phys. Rev. B* **77**, 144101 (2008).
28. Yamamoto, K. *et al*. Strong Optical Nonlinearity and its Ultrafast Response Associated with Electron Ferroelectricity in an Organic Conductor. *J. Soc. Phys. Japan* **77**, 074709 (2008).
29. Schrettle, F, Krohns, S., Lunkenheimer, P., Brabers, V.A.M. & Loidl, A. Relaxor ferroelectricity and the freezing of short-range polar order in magnetite. *Phys. Rev. B* **83**, 195109 (2011).
30. Shimizu, Y., Miyagawa, K., Kanoda, K., Maesato, M. & Saito, G. Spin liquid state in an organic Mott insulator with a triangular lattice. *Phys. Rev. Lett*. **91**, 107001 (2003).
31. Wang, H. H. *et al*. Phase Selectivity in the Simultaneous Synthesis of the T$_c$ = 12.8 K (0.3 kbar) Organic Superconductor κ-(BEDT-TTF)$_2$Cu[N(CN)$_2$]Cl or the Semiconductor (BEDT-TTF)Cu[N(CN)$_2$]$_2$. *Chem. Mater*. **4**, 247-249 (1992).

# Multiferroicity in an organic charge-transfer salt: Electric-dipole-driven magnetism

## Supplementary Information

P. Lunkenheimer[1]*, J. Müller[2], S. Krohns[1], F. Schrettle[1], A. Loidl[1], B. Hartmann[2], R. Rommel[2], M. de Souza[2,#], C. Hotta[3], J.A. Schlueter[4], M. Lang[2]

[1]Experimental Physics V, Center for Electronic Correlations and Magnetism, University of Augsburg, 86159 Augsburg, Germany
[2]Institute of Physics, Goethe-University Frankfurt, Max-von-Laue-Str. 1, 60438 Frankfurt(M), Germany
[3]Department of Physics, Faculty of Science, Kyoto Sangyo University, Kyoto 603-8555, Japan
[4]Materials Science Division, Argonne National Laboratory, Argonne, IL 60439, USA.
#Present address: Departamento de Física, IGCE, Unesp - Universidade Estadual Paulista, Caixa Postal 178, CEP 13500-970, Rio Claro (SP), Brazil
*e-mail: peter.lunkenheimer@physik.uni-augsburg.de

**Crystal structure and magnetic frustration**

Figure S1 shows the crystal structure of κ-(ET)$_2$Cu[N(CN)$_2$]Cl consisting of alternating layers of ET molecules (C$_{10}$S$_8$H$_8$) and insulating anion sheets. For materials with sufficiently large overlap of the molecular orbitals, the ET layer is conducting, providing a quasi-two-dimensional electronic band structure[1]. In the κ-type packing motif, the ET molecules form dimers, where adjacent dimers are arranged orthogonal to each other.

On the basis of the crystal structure, the hopping energies (transfer integrals $t$ and $t'$) of charges between adjacent molecules have been derived by *ab initio* calculations[2,3]. These studies yield a ratio of hopping energies for the holes between dimers, forming an anisotropic triangular lattice (see insets of Fig. 1 of main paper), of $t'/t \sim 0.44$ for κ-(ET)$_2$Cu[N(CN)$_2$]Cl and $t'/t \sim 0.83$ for κ-(ET)$_2$Cu$_2$(CN)$_3$ (ref. 2) Thus the degree of frustration $t'/t$ is distinctly smaller for κ-(ET)$_2$Cu[N(CN)$_2$]Cl as compared to κ-(ET)$_2$Cu$_2$(CN)$_3$.

For $T > T_{FE}$ the electrons reside on average between the dimers' molecules, forming an only weakly distorted triangular lattice. Upon entering the charge-ordered state below $T_{FE}$, however, the spins are located on a highly distorted triangular lattice, the symmetry of which is significantly reduced as compared to $T > T_{FE}$. In fact, when the calculation for κ-(ET)$_2$Cu$_2$(CN)$_3$ in ref. 10 of the main paper is applied to κ-

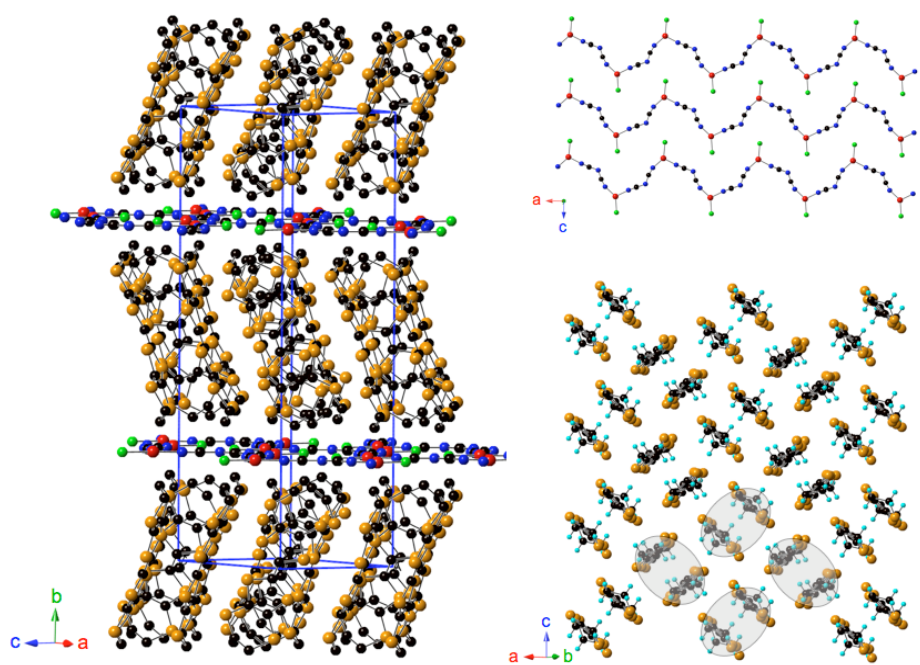

**Figure S1 | Crystal structure of κ-(ET)$_2$Cu[N(CN)$_2$]Cl.** Left: side view displaying the layered structure of ET molecules separated by thin insulating anion sheets. The unit cell is indicated by blue lines. Here, the hydrogen atoms are omitted for clarity. Right: top view of the anion layer (upper panel) and the ET layer (lower panel, selected dimers are highlighted by grey ellipses) showing the κ-type packing motif of the ET molecules. The direction perpendicular to the planes with the ET molecules is the crystallographic b axis. The anion layers are parallel to the ac plane. The polymeric anion chains are running along the a direction.

$(ET)_2Cu[N(CN)_2]Cl$, the value of $J'/J \propto (t'/t)^{1/2}$ in the charge-ordered state decreases to about 1/3 of that found in the dimer-Mott insulating state.

## Conductivity and Néel temperature

Figure S2 shows the temperature dependence of the conductivity $\sigma'$ of crystal 1, measured at various frequencies for $T < 50$ K. We found no significant frequency dependence of $\sigma'$ at $T > T_N \approx 27$ K. The frequency dependence observed at lower temperatures, revealed in Fig. S2, follows a power law, $\sigma' = \sigma_{dc} + \sigma_0 \nu^s$, with $s < 1$, typical for hopping conductivity[4]. However, at $\nu < 100$ Hz there is no detectable frequency dependence of $\sigma'$ down to the lowest temperature. Thus, the curve at 2.1 Hz shown in Fig. 1 of the main article provides a good estimate of the dc conductivity $\sigma_{dc}$. Below the CO transition close to $T_N$, which reduces the dc conductivity by two decades, obviously hopping becomes the dominant charge-transport process.

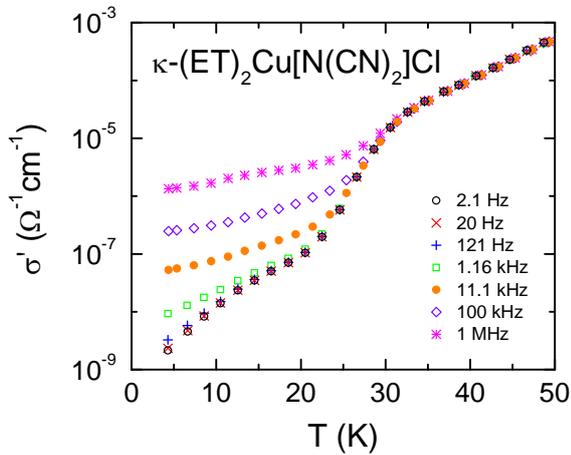

**Figure S2 | Temperature dependence of ac conductivity.** $\sigma'(T)$ of sample 1 as measured at various frequencies is shown for temperatures below 50 K.

Note that slightly different Néel temperatures have been reported, depending on the technique applied and the sample studied: While dc magnetization and $^1$H-NMR relaxation rate in the work by Miyagawa *et al.* yield a $T_N \sim 26 - 27$ K (ref. 5), the $^{13}$C-NMR studies by Smith *et al.* suggest a $T_N$ slightly in excess of 30 K (ref. 6). The latter value is consistent with anomalies observed in the resistivity by Ito *et al* (ref. 7). Lefebvre *et al.* report a Néel temperature of $T_N \sim 25$ K at ambient pressure deduced from a peak in the $^1$H-NMR relaxation rate[8].

Figure S3 shows $\sigma'(T)$ at 2.1 Hz in two representations for temperatures above the phase transition ($T > 30$ K). Obviously $\sigma_{dc}(T)$ neither follows a simple thermally activated behaviour nor a Mott's $T^{1/4}$-law[9] predicted by the variable-range-hopping model. One may speculate on the presence of two different activated regions[10] as indicated by the broken lines in Fig. S3a, but their significance should not be overemphasized.

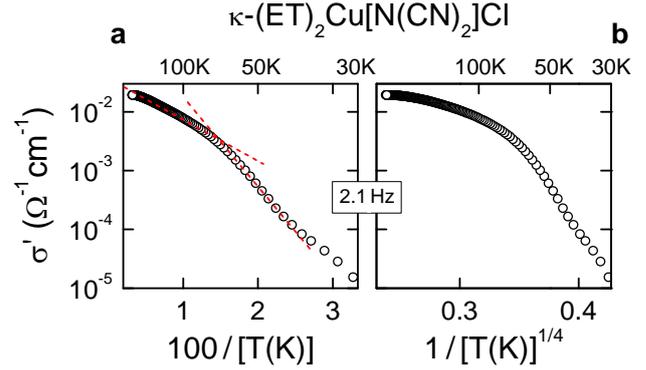

**Figure S3 | Temperature dependence of dc conductivity. a**, Conductivity of sample 1 at 2.1 Hz for $\mathbf{E}\|b$, plotted in an Arrhenius representation. The lines correspond to Arrhenius laws, $\exp(-\Delta E/(2k_B T))$ with $\Delta E = 26$ meV ($T > 70$ K) and 50 meV ($T < 70$ K). **b**, Same data as in a, plotted to achieve linear behaviour according to Mott's $T^{1/4}$-law.

## Dielectric response of sample 2

Figure S4 shows the dielectric constant $\varepsilon'(T)$ of sample 2 for various frequencies. In contrast to sample 1, for sample 2 measurements below 1 kHz were impeded by its higher conductivity. Otherwise, very similar behaviour to sample 1 is observed (cf. Fig. 2 of main article). The characteristic temperatures of sample 2 are somewhat higher (peak at $\sim 30$ K, $T_{CW} = 31$ K), than in sample 1. This may indicate different levels of defects, acting as dopands[11,12,13], also mirrored by its higher conductivity. The peaks are superposed to a somewhat lower background contribution than in sample 1 of about 60, mainly arising from stray-capacitance contributions. The dashed line in Fig. S4 demonstrates that the static dielectric constant $\varepsilon_s(T)$ is consistent with a Curie-Weiss behaviour with an additional background term, $\varepsilon_s = C/(T-T_{CW}) + \varepsilon_b$, leading to a Curie-Weiss temperature $T_{CW} = 31$ K $\approx T_{FE}$.

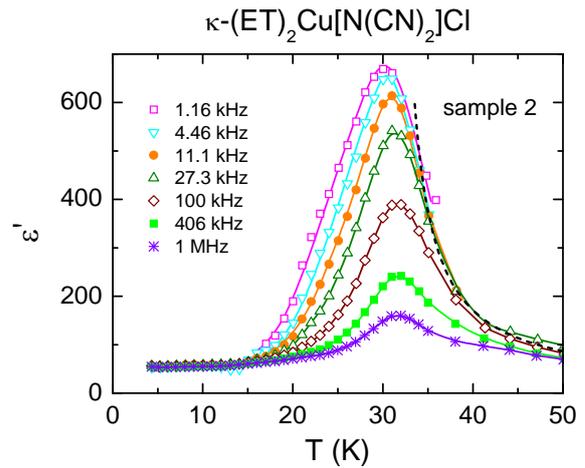

**Figure S4 | Temperature-dependence of the dielectric constant of sample 2.** $\varepsilon'(T)$ for sample 2, for selected frequencies and $\mathbf{E}\|b$. The solid lines are guides to the eyes. The dashed line demonstrates Curie-Weiss behaviour.

The absolute values of $\varepsilon'(T)$ of samples 1 and 2 are of similar order $\sim 10^2$ at 100 kHz. These numbers are distinctly smaller than those observed for the (TMTTF)$_2$X salts, where the dielectric anomalies reach values of order $10^5 - 10^6$ at 100 kHz (see, e.g., ref. 19 of the main paper). For the latter materials, the charge disproportionation could be directly observed through distinct line splittings in NMR[14] and infrared (IR)[15] spectroscopy. The comparatively small dielectric anomalies detected for the present material suggest that the corresponding features in NMR and IR spectroscopy are very small.

## Dielectric measurements in magnetic field

Figure S5 compares $\varepsilon'(T)$ at 56.2 kHz of sample 2 measured at a magnetic field of 9 T with **H**||*b* and at zero field. Obviously, the applied magnetic field has no effect on the dielectric constant. Similar results were obtained for fields up to 1.5 T, directed parallel to the planes. The inset shows field-dependent measurements performed at 30 K and several frequencies. Again no indications of any magnetocapacitive effects are found negating a spin-driven multiferroicity mechanism in κ-Cl.

There are some variations in the absolute values of the peak amplitude and position shown in Figs. S4 and S5. This may be ascribed to inner strains arising during cooling or strains produced from a small amount of "Apiezon N" grease applied to one side of the sample to ensure good thermal coupling to the sample mount of the PPMS. κ-Cl already becomes superconducting at 300 bar and a high sensitivity of the dielectric properties to pressure may be suspected. Pressure-dependent dielectric measurements are being prepared in order to address this point in detail.

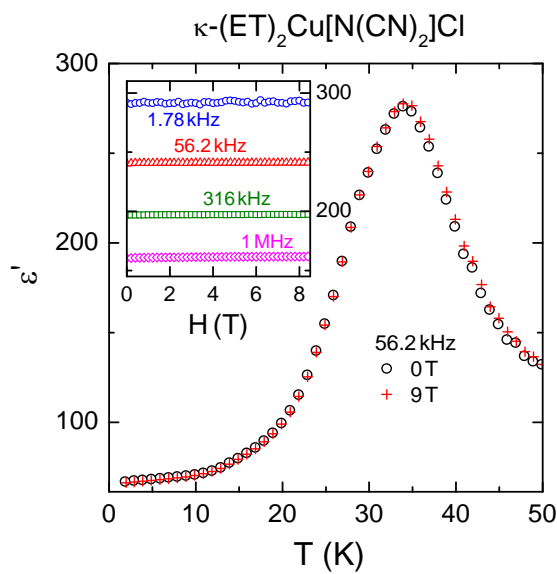

**Figure S5 | Magnetic field dependence of the dielectric constant.** The main frame presents $\varepsilon'(T)$ of sample 2, measured for **E**||*b* without and with a magnetic field of 9 T. The inset shows the dielectric constant at 30 K and various frequencies as a function of the magnetic field. For all measurements shown, the magnetic field was oriented perpendicular to the planes, i.e., **H**||*b*.

## Lattice effects

The observed ferroelectric transition implies that the inversion symmetry of the high-temperature structure is lost. The scenario suggested here, i.e., a combination of site- and bond-centred CO, naturally breaks the inversion symmetry. As pointed out in the main text, such an effect may be very small, beyond the resolution of structural investigations. However, as demonstrated for the (TMTTF)$_2$X salts[16], lattice effects accompanying the charge-order transition can be well resolved by means of thermal expansion measurements.

In fact, high-resolution thermal expansion measurements, performed both along the out-of plane and in-plane direction of κ-Cl[17], give clear evidence for lattice effects in the out-of-plane direction accompanying the CO transition, see Fig. S6. The results are qualitatively similar to those in the (TMTTF)$_2$X salts[16], where the charge modulation along the TMTTF stacks gives rise to displacements of the counterions.

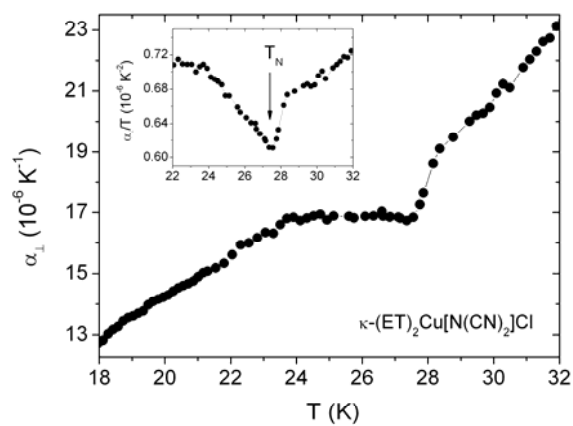

**Figure S6 | Thermal expansion coefficient perpendicular to the planes.** The main frame presents the linear coefficient of thermal expansion measured along the out-of-plane direction, $\alpha_\perp(T)$, of κ-Cl around the ferroelectric CO transition. The inset shows the same data in a plot $\alpha_\perp/T$ vs. *T* highlighting a distinct peak slightly above the reported Néel temperature of 27 K.

## References Supplementary Information

1. Toyota, N., Lang, M., & Müller, J. *Low-Dimensional Molecular Metals* (Springer, Berlin, 2007).
2. Kandpal, H.C. *et al.* Revision of model parameters for κ-type charge transfer salts: an ab initio study. *Phys. Rev. Lett.* **103**, 067004 (2009).
3. Nakamura, K. *et al.* Ab initio derivation of low-energy model for κ-ET type organic conductors. *J. Phys. Soc. Japan* **78**, 083710 (2009).
4. Elliott, S.R. Ac conduction in amorphous semiconductors. *Adv. Phys.* **36**, 135-215 (1987).
5. Miyagawa, K. *et al.* Antiferromagnetic ordering and spin structure in the organic conductor κ-(ET)$_2$Cu[N(CN)$_2$]Cl. *Phys. Rev. Lett.* **75**, 1174-1177 (1995).
6. Smith, D. F. *et al.* Dzialoshinskii-Moriya interaction in the organic superconductor κ-(BEDT-TTF)$_2$Cu[N(CN)$_2$]Cl. *Phys. Rev. B* **68**, 024512 (2003).
7. Ito, H. *et al.* Metal-Nonmetal transition and superconductivity localization in the two-dimensional conductor κ-(BEDT-TTF)$_2$Cu[N(CN)$_2$]Cl under pressure. *J. Phys. Soc. Japan* **66**, 2987-2993 (1996).

8. Lefebvre, S. *et al.* Mott transition, antiferromagnetism and unconventional superconductivity in layered organic superconductors. *Phys. Rev. Lett.* **85**, 5420-5423 (2000).
9. Mott, N. F. & Davis, E. A. *Electronic processes in non-crystalline materials* (Clarendon Press, Oxford, 1979).
10. Williams, J. M. *et al.* From semiconductor-semiconductor transition (42 K) to the highest-$T_c$ organic superconductor, κ-(ET)$_2$Cu[N(CN)$_2$]Cl ($T_c$ = 12.5 K). *Inorg. Chem.* **29**, 3272-3274 (1990).
11. Analytis, J. G. *et al.* Effect of irradiation-induced disorder in the conductivity and critical temperature of the organic superconductor κ-(BEDT-TTF)$_2$Cu(NCS)$_2$. *Phys. Rev. Lett.* **96**, 177002 (2006).
12. Sasaki, T. *et al.* X-ray irradiation-induced carrier doping effects in organic dimer-Mott insulators. *J. Phys. Soc. Japan* **76**, 123701 (2007).
13. Komatsu, T. Realization of superconductivity at ambient pressure by band-filling control in κ-(BEDT-TTF)$_2$Cu$_2$CN$_3$. *J. Phys. Soc. Japan* **65**, 1340 (1996).
14. Chow, D.S. *et al*. Charge Ordering in the TMTTF Family of Molecular Conductors. *Phys. Rev. Lett.* **85**, 1698 (2000).
15. Dumm, M. *et al.* Mid-infrared response of charge-ordered quasi-1D organic conductors (TMTTF)$_2$X. *J. Phys. IV (France)* **131**, 55 (2005).
16. De Souza, M. *et al.* Evidence for lattice effects at the charge-ordering transition in (TMTTF)$_2$X. *Phys. Rev. Lett.* **101**, 216403 (2008).
17. J. Müller, PhD Thesis, TU Dresden (2002).